# Optimizing Post-Cancer Treatment Prognosis: A Study of Machine Learning and Ensemble Techniques

Joyee Chakraborty[1*†], Mazahrul Islam Tohin[1†], Danbir Rashid[1, 2†], Tanjil Ahmed Tanmoy[1†], Md. Jehadul Islam Mony[1†]

[1]Department of Computer Science and Engineering, Leading University, Ragibnagar, Sylhet 3112, Bangladesh.
[2]DeepNet Research and Development Lab, Sylhet 3100, Bangladesh.

*Corresponding author(s). E-mail(s): joyeeck04@gmail.com;
Contributing authors: tohin214@gmail.com; danbirrashid54@gmail.com; tanziltanmoy21@gmail.com; mony_cse@lus.ac.bd;
[†]These authors contributed equally to this work.

## Abstract

The aim is to create a method for accurately estimating the duration of post-cancer treatment, particularly focused on chemotherapy, to optimize patient care and recovery. This initiative seeks to improve the effectiveness of cancer treatment, emphasizing the significance of each patient's journey and well-being. Our focus is to provide patients with valuable insight into their treatment timeline because we deeply believe that every life matters. We combined medical expertise with smart technology to create a model that accurately predicted each patient's treatment timeline. By using machine learning, we personalized predictions based on individual patient details which were collected from a regional government hospital named **Sylhet M.A.G. Osmani Medical College & Hospital, Sylhet, Bangladesh**, improving cancer care effectively. We tackled the challenge by employing around 13 machine learning algorithms and analyzing 15 distinct features, including Logistic Regression, Support Vector Machine, Decision Tree, Random Forest, etc we obtained a refined precision in predicting cancer patient's treatment durations. Furthermore, we utilized ensemble techniques to reinforce the accuracy of our methods. Notably, our study revealed that our majority voting ensemble classifier displayed exceptional performance, achieving 77% accuracy, with LightGBM and Random Forest closely following at approximately 76% accuracy. Our research unveiled the inherent complexities of cancer datasets, as seen in the Decision Tree's 59% accuracy. This emphasizes the need for improved algorithms to better predict outcomes and enhance patient care. Our comparison with other methods confirmed our promising accuracy rates, showing the potential impact of our approach in improving cancer treatment strategies. This study marks a significant step forward in optimizing post-cancer treatment prognosis using machine learning and ensemble techniques.

# 1 Introduction

In current times, despite having modern medical technology, Cancer is a disease that still causes death [1]. It is a large group of diseases that can start in almost any organ or tissue of the body. It happens when cells in a body start to grow abnormally and uncontrollably. Cells start to grow beyond their boundaries which invade adjoining parts of the body and/or spread to other organs [2]. According to the World Health Organization (WHO), Cancer is the second leading cause of human death around the globe. In the year of 2018, 1 in 6 deaths was caused by cancer[3], which is more than AIDS, tuberculosis, and malaria combined. Each year, approximately 400,000 children develop cancer [4]. The most common causes of cancer death in 2020 were lung, breast, colon and rectum, liver and stomach[5]. It became a major burden of disease worldwide. Each year, tens of millions of people are diagnosed with cancer around the world, and more than half of the patients eventually die from it[6]. Cancer can occur from a person's genetic factors and external agents such as physical carcinogens eg. ultraviolet and ionizing radiation, chemical carcinogens eg. asbestos, tobacco smoke, and alcohol, etc., or biological carcinogens eg. viruses, bacteria, or parasites. It becomes more deadly for older people as their cellular repair mechanisms become less effective[7]. Furthermore, some chronic infections also cause cancer which is a big issue for low and middle-income countries. Approximately 13% of cancers diagnosed in 2018 globally were attributed to carcinogenic infections, including Helicobacter pylori, human papillomavirus (HPV), hepatitis B virus, hepatitis C virus, and Epstein-Barr virus. Between 30 and 50% of cancers can be prevented by avoiding risk factors and the chances for cure becomes high if it is detected in an early stage and appropriate treatment is done.

Cancer detection in the early stage is very important. It can be cured if detected early. Traditionally, cancer detection depends on various diagnostic tests such as Biopsies, Blood tests, PEP tests, Imaging scans, MRI, PET scans, Endoscopy, Sigmoidoscopy, Mammogram, Colonoscopy, Physical examination, Bronchoscopy, Sputum cytology, Ultrasound, Genetic tests, Urinalysis, Circulating tumor cell tests, PSA test etc by the medical professional interpreting[8]. The human brain and eye can only process a limited amount of information, making it very challenging for early cancer detection. Nowadays Technology such as Machine learning has our lives greatly eased. Machine learning is a very powerful tool for cancer detection, offering a transformative approach to diagnosing the disease [9]. By using a vast amount of patient data, ML algorithms can identify nice patterns and irregularities indicative of cancerous growths with remarkable accuracy. From analyzing medical imaging such as MRI and CT scans to parsing genetic markers and biomarkers in blood samples, ML algorithms excel in recognizing early signs of malignancy that may avoid human detection. Furthermore, these algorithms continuously learn and improve over time refine their ability, and enhance diagnostic precision and efficiency. Also using ML algorithms for cancer detection reduces healthcare costs by streamlining the diagnostic process and minimizing unnecessary procedures and tests[10].

Throughout our research, we used a wide range of machine learning classifiers, such as the Naive Bayes approach, Artificial Neural Networks, K-Nearest Neighbors strategy, Random Forest algorithm, Support Vector Machine technique, Logistic Regression analysis, and Decision Tree classifier. These classifiers were trained in a supervised learning environment, where each model was given access to a predetermined amount of data to work with. A variety of evaluation criteria, including accuracy, precision, recall, and the F-measure, were used to assess each classifier's effectiveness. Versions 1.4.0 and 3.12.2 of the Python programming language, along with the Pandas package for back-end processing, served as the foundation for our investigation. The dataset used in the study was subjected to a thorough preparation step during which any missing values were found and removed. Following this, we ensured consistency in the magnitude of the data points by applying a normalization approach to the remaining data. In the latter portion of our paper, we go into great depth on the findings of this investigation. Through the discussion, readers will gain a clear knowledge of the dependability and implications of the various machine learning algorithms used in the field of oncological prediction. Closing remarks that summarize the main points of the work done and its possible significance round out the essay. Furthermore, we provide an outlook on how our research can develop going forward, outlining possible directions and approaches that could improve the predicted accuracy and dependability for clinical applications in the future.

The following structure applies to the remaining portion of the paper: The papers literature review portion is included in Section 2. Research goals are covered in depth in Section 3. Part 4 provides a detailed explanation of the methodologies. The proposed methods results are shown in Section 5. To sum up, the manuscript is concluded in Section 6. In today's world, cancer is a widespread challenge that has affected many people, including our own families. We have personally experienced the challenges and emotional impact of this disease. It is this deep, personal connection that fuels our motivation to make a meaningful difference in cancer care. Our journey begins with the profound understanding that each person battling cancer deserves personalized care and accurate treatment timelines. We are driven by the belief that every life matters and that each patient's well-being is of utmost importance. With this conviction at heart, we set out to create a new approach to cancer treatment, one that combines medical expertise with smart technology. With a focus on machine learning techniques and various features, we are dedicated to achieving a high level of accuracy in predicting treatment durations. Ultimately, our motivation is to make a real difference in cancer treatment and to bring hope and healing to those in need. This motivation drives our dedication and serves as the guiding force for our research endeavors. Our goal is to enhance patient care and recovery by bringing precision and personalization to the forefront of cancer treatment.

## 2  Background Study

In our thesis paper, our objectives are to examine the necessity of specialized treatments in cancer care within Sylhet, Bangladesh. We aim to explore the potential benefits of advanced technology, specifically machine learning algorithms, in improving treatment outcomes. Our focus includes understanding how the user interface of such systems can facilitate treatment decisions for healthcare providers. Additionally, we intend to investigate the impact of machine learning on cancer treatment decision-making processes. Through our research, we aim to contribute insights and recommendations to enhance cancer care practices in Sylhet.

Sharma et al.'s [11] study uses SVM and random forest methods to classify bone cancer with a 92% accuracy using a dataset of 105 X-ray images from different sources. The study's methodology includes an image processing procedure that enhances the quality of input data to increase model performance. The modest size of the dataset, however, raises concerns about the generalizability of the suggested approach. Despite this, the work represents a significant advancement in the field and shows how machine learning might enhance the identification of bone cancer in medical diagnostics. To ensure broader use in clinical settings, future research should primarily address the limits of the dataset.

Gupta et al.'s [12] study addresses the important problem of colon cancer prognosis using text-based datasets. The research uses machine learning models, such as Random Forest, SVM, MLP, AdaBoost, and Logistic Regression, to forecast disease-free survival and determine tumor stages. It gained 89% accuracy rate with the Random Forest classifier.The dataset provides a solid foundation for the predictive models, totaling 4021 cases. However, the literature points to a flaw in the insufficient explanation of the methods employed, emphasizing the need for more accurate documentation in follow-up research to enhance repeatability and understanding.

Montazeri et al.'s [13] study was to predict breast cancer survival using machine learning algorithms. The study employed text-based datasets and employed features including NB, TRF, 1NN, AD, SVM, RBFN, and MLP for classification. The qualities were reported appropriately, however the study's methodology was found imprecise. The dataset, which included information from 900 patients, offered significant new insights into the prediction of breast cancer survival rates. Taking everything into account, this study showed how machine learning might enhance models for predicting the fate of breast cancer.

Alabia et al.'s [14] study was to apply supervised machine learning classification techniques to predict early oral tongue cancer. The research provided a detailed examination and comparison of four well-known algorithms: Decision Forest (DF), Boosted Decision Tree (BDT), Naive Bayes (NB), and Support Vector Machine (SVM), using text-based datasets. The transparency of the process was improved by the characteristics utilized in the study having thorough documentation. The study's focus on artificial intelligence in predicting oral tongue cancer demonstrates the promise of machine learning in medical applications. All things considered, the study provides valuable information regarding the connection between healthcare and disease prediction algorithms.

Islam et al.'s [15] study presents a promising approach to breast cancer prediction using Support Vector Machine (SVM) and K-Nearest Neighbors (K-NN) algorithms, achieving high accuracies of 98.57% for SVM and 97.14% for K-NN in the testing phase. However, limitations include the unbalanced nature of the dataset, lack of comparison with other models, and testing on a single dataset. To improve, the researchers could validate the models on more diverse and larger datasets, employ techniques to address unbalanced data, incorporate additional features from medical imaging and records, and conduct cross-institutional validations to enhance the model's generalizability.

Diamant et al.'s study [16] focuses on using convolutional neural networks (CNNs) to predict treatment outcomes of head and neck (H&N) cancer patients based solely on pre-treatment CT images, aiming to aid in patient risk stratification and treatment selection. The research found success in developing an end-to-end CNN framework capable of recognizing radiomic features with proven predictive power, outperforming traditional radiomics. However, potential limitations include the effectiveness of transfer learning for recognizing radiomic features and the model's reliance on the central tumor slice. To improve, the authors could integrate 3-dimensional information and PET images, explore additional visualization tools, simplify feature engineering requirements, and enhance transfer learning approaches to bolster the model's performance.

Chu et al.'s study [17] aims to use CNNs to predict treatment outcomes for head and neck cancer patients based solely on pre-treatment CT images, aiding in patient risk stratification and treatment selection. The research successfully develops an end-to-end CNN framework capable of recognizing radiomic features and predicting oncological outcomes. Potential limitations include the transfer learning approach and reliance on the central tumor slice for training and evaluation. To enhance the study, the authors could integrate 3D information and PET images, explore additional visualization tools, and improve transfer learning and combination approaches to bolster the model's performance.

Zhu et al.'s paper [18] examines the use of deep learning in cancer prognosis prediction, highlighting its potential to enhance clinical decision-making by utilizing various health data types. The study reviews the application of deep learning models, including neural network architectures, in handling multi-omics data, clinical information, and imaging for nuanced survival estimation. It acknowledges the challenge of limited patient data for training, potentially impacting model reliability and performance due to overfitting. To address these limitations, the paper suggests the exploration of techniques for small sample sizes, the management of imbalanced data, feature extraction enhancement, and the development of secure data infrastructure. Furthermore, it emphasizes the need for interdisciplinary collaboration to advance the field's application.

Vahini et al.'s article [19] offers valuable insights into early liver disease detection through machine learning applied to CT scans. By leveraging morphological operations and computed tomography to highlight tumor regions, the research presents a significant contribution to improving patient outcomes through prompt identification. However, the article lacks transparency regarding the specific machine learning algorithms utilized, and the model's performance limitations with single masses of tumors indicate a need for refining the approach to address multiple liver tumors. Moving forward, refining the methodology to inclusively evaluate and validate multiple tumors, in addition to providing greater clarity on the employed machine learning algorithms, could significantly enhance the study's impact on early liver disease detection and treatment efficacy.

Asri et al.'s paper [20] investigates the application of machine learning algorithms in classifying breast cancer data, focusing on the comparison of Support Vector Machine, Decision Tree, Naive Bayes, and k Nearest Neighbors. It underscores the significance of accurate classification in the medical domain and highlights the potential of data mining techniques in improving healthcare outcomes. Through experimental analysis using the Wisconsin Breast Cancer dataset, the study demonstrates that SVM achieves the highest accuracy of 97.13% with the lowest error rate. One concern with this paper is that it doesn't talk much about how to pick the best features from the data or how to simplify the data to make it easier for the algorithms to work with.

Deepshikha et al.'s study [21] highlighted the significance of machine learning in medical imaging for cancer detection, particularly focusing on bone cancer. It discusses various studies that utilize machine learning algorithms such as random forests, support vector machines, decision trees, genetic algorithms, and swarm intelligence for classification tasks. These studies employ diverse methodologies including feature extraction, segmentation, thresholding, and clustering techniques on medical imaging data like CT scans and MRI images. While acknowledging the successes achieved, the review also underscores challenges such as the need for standardized benchmarking, integration of molecular signatures with imaging data, and improving accuracy and predictive power.

Usha et al.'s paper [22] reviews methods for detecting and classifying skin cancer, including preprocessing, segmentation, and classification techniques. Studies use various approaches such as color-based analysis, texture analysis, and machine learning algorithms like SVM and naive Bayes. While promising, these methods face challenges such as dataset specificity and limited generalization. Future research could focus on hybrid approaches and advanced machine learning to improve accuracy and efficiency.

Islam et al.'s paper [23] compares five machine-learning algorithms for breast cancer detection. These are Support Vector Machine, K-Nearest Neighbors, Random Forests, Artificial Neural Networks, and Logistic Regression. It aims to develop an automated system for early breast cancer detection. By analyzing the dataset, the study evaluates algorithm performance using metrics like accuracy and sensitivity. But it doesn't talk much about understanding why these methods work or if they will work well in different situations. More research is needed to understand these things better.

Salmi et al.'s paper [24] highlights the importance of early detection due to the lack of obvious symptoms in the early stages of colon cancer. The study utilizes machine learning techniques to classify patients based on their cancer status. By proposing the Naive Bayes Classifier model, the paper demonstrates high accuracy in classifying patients as either suffering from colon cancer or not. The method's simplicity and ability to handle large datasets make it a promising tool for cancer classification. However, a drawback mentioned is the assumption of independence between attributes, which may reduce accuracy in cases where attributes are related.

Although conventional methods are still being used in the medical field, machine learning techniques have shown itself to be reliable enough for use in this setting. According to recent research, these techniques are precise enough to be applied to a variety of tasks, such as classification, prediction, and decision-making in many medical fields. Our study aims to demonstrate how real-world medical data may be processed to create a model that can be used to calculate the time duration for treatment of various cancer types.

## 3 Proposed Methodology

### 3.1 Data Acquisition

The Cancer datasets were collected from OSMANI MEDICAL COLLEGE HOSPITAL, Sylhet, Bangladesh, under the ethical considerations of the Leading University Ethical Committee. The datasets contain thousands of primary data. Each primary dataset contains three to four pages of information about one cancer patient, including patient demographics, medical history, treatment history, clinical notes, chemo history, various test results, and reports. One of our toughest tasks was to collect this 'data goldmine' and convert it into an Excel file. Due to the versatility of these datasets, they can be used for various studies. The key features are explained in Table 1.

### 3.2 Data Preprocessing

Data Preprocessing is the most important task for any data science project, as the result depends on how the datasets were preprocessed[25]. If your dataset carries outliers or anomalies at training, it's pretty sure that you'll end up with a less accurate and higher-biased model. Here are some steps we took in preprocessing our datasets:

#### 3.2.1 Duplicates

Datasets with duplicates can skew the analysis and lead to overfitting[26]. Therefore, it is necessary to remove them to minimize bias in the datasets. Our datasets included 33 duplicate rows, which were simply dropped to ensure the uniqueness of the datasets.

#### 3.2.2 Handling missing values

The toughest task in dealing with null values is deciding whether it's okay to drop them or how to replace them. Since our datasets had a bit more missing values, we focused on the latter question. We applied some conventional techniques to replace them[27]. One of our key features 'weight' had almost 10% null values. To handle them, we used a RandomForest Classifier. Initially, we selected four other predictor features for the 'weight' column based on the correlation matrix and predicted the 'weight' column for it's valid data. Our model achieved an accuracy of 100% and 99% for the training and testing data, respectively, using a 7:3 ratio. Finally, We replaced the null values in the 'weight' column by using this model.

| No. | Features | Data Types | Description |
| --- | --- | --- | --- |
| 1 | GENDER | Nominal | Indicate the patient's gender: 'm' for male; 'f' for female. |
| 2 | AGE | Numerical | Indicate the age of patients, Some instances are - 23, 34, 57, etc; these are the age (e.g., 23 years) of patients in integer format. |
| 3 | MARITAL STATUS | Nominal | Marital status of patients: 'married', 'unmarried'. |
| 4 | DISTRICT | Nominal | District of the patients: 'sylhet', 'sunamganj', 'moulvibazar', 'habiganj'. |
| 5 | OPERATION HISTORY | Nominal | Do patients have any operation history: 'yes', 'no'. |
| 6 | WEIGHT | Numerical | Current weight of patients in numeric form; e.g., 34, 56, 44, 86, etc; these values indicate the patient's weight in kilograms. |
| 7 | DIET | Nominal | What kind of diet does the patients follow currently: 'normal', 'standard', 'soft & liquid', 'diabetic'. |
| 8 | CANCER TYPE | Nominal | Indicates the specific type or subtype of cancer that a person has been diagnosed with. |
| 9 | TOTAL CHEMO CYCLE | Numerical | The number of chemo cycles prepared for the patients for his chemotherapy by the specialist or doctors. It's instances are: 0, 1, 2, 3, 4, 5, 6, 7, 8 (in numeric). |
| 10 | PLANNING CYCLE | Numerical | It indicates that currently patients are taking at which cycle chemo injections. It's instances are: 0, 1, 2, 3, 4, 5, 6, 7, 8 (in numeric). |
| 11 | CHEMO | Nominal | It indicates the chemo injection which will be provided in a particular cycle, and it can be multiple chemo injections at a cycle(e.g. 'cisplatin', 'ifosfamide', 'doxorubicin') or maybe a single injection at a particular cycle(e.g. 'oxaliplatin'). |
| 12 | TOTAL CHEMO WEEKS | Numerical | It means the number of weeks it takes to complete a chemo cycle, instances are: 1, 2, 3, 4, 5 (in numeric). |
| 13 | TREATMENT PERIOD | Numerical | This is nothing but the amount of time a patient takes or will take treatment on a particular cycle. |

**Table 1** Features of Patients in a Particular Cycle

### 3.2.3 Outliers Detection

Outliers are data points that may be errors, or they may represent variations in the data. We got some male patients who had breast cancer, although studies [28] show that 1 of 100 breast patients is male. Some males had ovary cancer, and some females had testicular cancer. for treatment, patients got a higher level of planning cycle than the total possible cycle. these are anomalies or outliers. One of the assumptions is that they may have occurred due to typical errors in creating the datasets. We have cleared up all of these errors.

### 3.2.4 Dimensionality Reduction

It's important to reduce unuseful columns by conducting analysis, such as using a correlation matrix. Sometimes, for lack of data, we classify some classes into a similar group as the 'other' category. It not only solves the overfitting problem but also enhances the efficiency and effectiveness of the model[29]. Principal Component Analysis (PCA), a classic technique of dimensionality reduction[30], was used in our model.

### 3.2.5 Feature Engineering

At this stage, we apply label encoding and one-hot encoding to the categorical features based on their categories, types, and impacts. According to data visualization, we identified some features that showed exactly a bell-shaped distribution. Therefore, we picked normalization instead of standardization to scale them[31].

## 3.3 Feature Analysis

Data analysis is important for visualizing the data, understanding the relationships between different features, and making decisions for the next steps[32]. Matplotlib and Seaborn are two libraries used during the analysis of different features. One of the major analyses is explained below-

### 3.3.1 Correlation Matrix

A correlation matrix is a technique used to analyze the relationships between different features, particularly when the features are in numeric format. Recently, it has become one of the most widely used techniques for finding the best features related to the target column. And at some point you need to remove the less correlated features for the problem of high dimensionality[33]. In our work, We built a correlation matrix to check the relationship of different features with respect to the target column.We found out some best features and some less correlated features.The less correlated features are dropped from the datasets to reduce the dimensionality and potential biases. The heatmap depicts the correlation matrix in Figure 1.

## 3.4 Data Splitting

We have prepared our datasets into a suitable numeric format that can be used by any machine learning model. Its time to split the data into training and testing sets. We allocated 80% of the data for training and the remaining 20% for testing. Although we considered other combinations (such as a 7:3 ratio or an 85:15 ratio), the most versatile and preferable version is the 80:20 ratio[34].

## 3.5 Model Training

Model training in machine learning involves using algorithms to learn patterns in data for making predictions[35]. Techniques range from traditional methods like Logistic Regression and Naive Bayes to advanced approaches like Support Vector Machines,

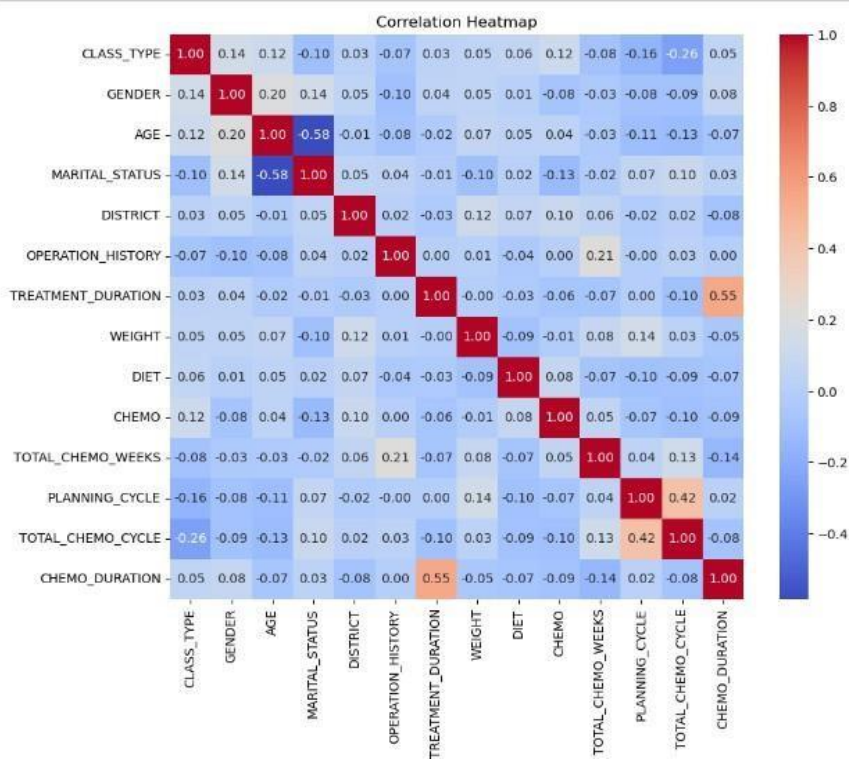

**Fig. 1** Corelation Matrix

Random Forest, and Neural Networks. Each model has hyperparameters that require tuning for optimal performance[36], assessed by metrics such as accuracy, precision, recall, and F1-score. Cross-validation ensures robustness, while hyperparameter tuning refines model performance.

### 3.5.1 Logistic Regression

Logistic Regression is nothing but the linear combination of input features with a special property of predicting discrete classes[37]. The linear combination represents the weighted sum of input features, often referred to as the *logit*. Given that the input features are $[x_1, x_2, \ldots, x_n]$ and corresponding weights are $[\vartheta_1, \vartheta_2, \ldots, \vartheta_n]$ with intercept $\vartheta_0$. then the logit will be,

$$\vartheta^T x = \vartheta_0 + \vartheta_1 x_1 + \ldots + \vartheta_n x_n$$

The special property, i.e., - logistic function $h_\vartheta(x)$, is used to transform the 'logit' into a probability score, which allows us to interpret the results as the likelihood of belonging to a specific class. The sigmoid function:

$$h_\vartheta(x) = \frac{1}{1 + e^{-\vartheta^T x}}$$

To predict multiclass, we apply both 'One vs Rest' and 'Multinomial' properties with other hyperparameters For example - solver as a [ 'liblinear', 'lbfgs', 'newton-cg'], and different C values, among them, it's found out that 'One vs Rest' with 'newton-cg' solver utilizes the most efficiency and accuracy of the model.

### 3.5.2 Support Vector Machine

When it's about maximizing the margin of a particular scatter, the first thing that comes to mind is the SVM algorithm[38]. The Support Vector Machine is a widely used supervised learning algorithm that chooses the hyperplane space in such a way that it closely represents the underlying function in the target space [39]. In this proposed work, the SVM model was tuned by picking different values and properties of 'gamma', 'C' and 'kernels' parameters and finding out which of the parameters results more accurately.

### 3.5.3 Naive Bayes

Naive Bayes is a probabilistic classifier that works based on the conditional probability between the predictor and target class, with the principle of Bayes Theorem[40]. One of the variants, Multinomial NB introduced in our work due to the characteristics of the datasets.

### 3.5.4 Decision Tree

A tree consists of nodes, branches, and leaves.[41] In Decision Tree, Leaf nodes represent the prediction while the internal nodes represent the decision of whether it is required further branching or which way to traverse or predict the target value based on any particular predictor column. The purity of the decision at any internal nodes is based on how you tune the hyperparameters.

### 3.5.5 Random Forest

An ensemble bagging version with multiple weak learners, that's Random Forest. The weak learners are the Decision Trees which are trained on random subsets of the datasets.[42] And finally, aggregate the result using majority voting(for classification problem) or average(for regression problem) of all trees. In our work, We tuned the RF classifiers by checking which of the combination of values is suitable for parameters like, 'the number of estimators', 'the maximum depth of the trees', ' whether it is gini or entropy criterion', etc.

### 3.5.6 KNN

The K-Nearest Neighbors Algorithm consists of finding the distance between the label and unlabeled data[43]. The distance must be sorted in ascending order, starting with the first 'K' instance of them chosen and checking the majority voted class, that's the predicted class. By definition, it's guaranteed that 'K' is an odd number because at some point, even 'K' can produce a tie in voting, e.g.- A binary classification with k = 4 for the KNN model achieves at predicting 2 label data of class - '0' and 2 for class -'1', that's called a tie. In such a situation, the algorithm will be confused about choosing any class[44]. By tuning, the model can be improved due to the impact of different hyperparameters on the characteristics of datasets.

### 3.5.7 Gradient Boosting Machine

GMB operates sequentially by training weak learners typically the trees. Each subsequent tree is trained to correct the errors made by the previous tree[45]. Finally, combine the results of all trees to predict the target column which is actually better than any individual model.

### 3.5.8 AdaBoost

Adaptive Boosting technique added the concept of weighted errors to iteratively improve the performance of weak learners[46]. With the number of data points is N, the initial weight of each data points, $w = \frac{1}{N}$. Generally the misclassified data of previous base learners assigned with higher weight and accurately classified data assigned with a way lower weight. Update weight measures as, For misclassification, $w = w_{old} + e^{performance}$ For correct classification, $w = w_{old} + e^{-performance}$ The measure of performance after each iteration, performance = $\frac{1}{2} \log_e \frac{1-weightedErrors}{weightedErrors}$ After completing the number of iterations with a particular criterion the model predicts the outcome of the target column.

### 3.5.9 XGBoost

Extreme Gradient Boosting is one of the optimized and powerful boosting techniques that especially works for the performance and efficiency of the model with the help of combinations of base learners. [47] The hyperparameter tuning matters for its accuracy, the tuned parameters in our model are the number of estimators, learning rate, maximum depth of the trees, and minimum child weight, as they had the most variance, after tuning them, there were no such variance that can affect the accuracy of the model.

### 3.5.10 LightBoost

LightBoost stands for Light Gradient Boosting Machine. LightBoosting supports categorical features without encoding, it is also known for its scalability and memory management.[48] The parameters used to tune the LightGBM classifiers are the number of iterations or estimators, the learning rate, the maximum depth of the base learner of the trees, the maximum number of leaves in each tree, and the minimum number of data required in each leaf. As the number of leaves can grow exponentially, it is necessary to think about complexity, A good practice is to set the number of leaves to less than $2^{max\ depth}$.

### 3.5.11 CatBoost

When your datasets burst with categorical features and you want to use some boosting technique, you are welcome to apply cat boosting, which is specially made for dealing with categorical features. It does not require preprocessing like one-hot encoding of categorical variables, as it applies its built-in encoding techniques.[49] Recently, it has become one of the most used algorithms in machine learning.[8] The tuning of its parameters varies the accuracy, we tuned the model with the best parameters like the depth of the tree, the number of iterations, the learning rate, etc.

### 3.5.12 ANN

In the human brain, the processing of information happens via the electrochemical signals between billions of neurons
through synapses.[50] The same techniques are used in Neural networks to process information and predict the target value. The neurons are the activation functions, and the synapses are the connections between them. ANNs are structured into layers, the first and last layers are the input and output layers and all intermediate layers are called hidden layers. The hidden layers improved the ability of the networks to learn different complex patterns of the datasets.[51] In our study, We applied 1-5 hidden layers. The neurons, by which I mean applied activation functions are - 'tanh', 'identity', 'logistic', and 'relu'. We also tune other parameters such as 'a number of iterations', 'learning rate', 'alpha', and 'solver' from which we consider different ranges of values based on their impact on the model performance.

### 3.5.13 Voting Classifiers:

Voting Classifier is an ensemble technique that differs from Bagging, Boosting, and even Stacking ensemble techniques instead it is a customized ensemble technique with multiple chosen base learners[52]. It primarily applies the majority voting technique among all base learners, especially when voting is hard, which is common for classification problems. However, it averages the results of base learners when voting is soft, which is often the case for regression problems.

## 3.6 Evaluation

The built models, such as Logistic Regression, Support Vector Machine, Decision Tree, Random Forest, Naive Bayes, KNN, Gradient Boosting Machine, XGBoost, AdaBoost, LightGBM, CatBoost, and ANN, are verified using different evaluation metrics and cross-validation techniques.

The metrics used include the confusion matrix, accuracy, precision, recall, and F1-score. For cross-validation, a Stratified K-Fold was applied, as it ensures that all types of data points are represented in each fold, unlike traditional K-Fold. Regarding hyper-parameter tuning, RandomizedSearchCV was applied to select the best parameters for each model.

# 4 Result and Analysis

## 4.1 Performance of Models with All Features

In our analysis, we explored a diverse range of machine learning algorithms, including logistic regression, support vector machines (SVM), decision trees, random forest, k nearest neighbors (KNN), naive Bayes, gradient boosting, XGBoost, AdaBoost, LightGBM, CatBoost, and artificial neural networks (ANN). Each algorithm offers unique strengths and characteristics. Logistic regression provides a simple yet interpretable model for binary classification. SVM is effective for high dimensional data and can handle non linear relationships through kernel functions. Decision trees offer intuitive decision- making processes, while random forest aggregates multiple trees for improved accuracy and resilience to overfitting. KNN relies on instance-based learning and is suitable for non-parametric classification tasks. Naive Bayes models are computationally efficient and work well with categorical features. Gradient boosting methods sequentially improve model performance by minimizing errors. XGBoost, AdaBoost, LightGBM, and CatBoost are advanced boosting algorithms with optimizations for efficiency and accuracy. Finally, artificial neural networks, inspired by the human brain, are versatile models capable of learning complex patterns but require substantial computational resources and data. By exploring these algorithms, we gained insights into their suitability for various tasks and dataset characteristics.

| Algorithm | Accuracy | Precision | Recall | F1 Score |
| --- | --- | --- | --- | --- |
| Logistic Regression | 71% | 63% | 71% | 67% |
| SVM | 71% | 57% | 71% | 63% |
| Decision Tree | 57% | 62% | 57% | 59% |
| Random Forest | 70% | 66% | 70% | 67% |
| KNN | 68% | 60% | 68% | 63% |
| Naive Bayes | 64% | 62% | 64% | 62% |
| Gradient Boosting | 72% | 74% | 72% | 71% |
| XGBoost | 70% | 68% | 70% | 68% |
| AdaBoost | 41% | 69% | 41% | 48% |
| LightGBM | 71% | 72% | 71% | 70% |
| CatBoost | 72% | 65% | 72% | 68% |
| ANN | 59% | 62% | 59% | 60% |

**Table 2** Model Performance Metrics

It appears that the classifiers evaluated have varying levels of performance according to the metrics of accuracy, precision, recall, and F1 score. The Gradient Boosting Classifier demonstrates the highest effectiveness among the listed algorithms, with the highest accuracy of 72%, along with the highest precision at 74%, a recall of 72%, and an F1 score of 71%. Following Gradient Boosting, both the Logistic Regression and LightGBM perform similarly in terms of accuracy, each achieving 71%. However, LightGBM shows a slightly better precision of 72% compared to Logistic Regression's 63%. Recall for both algorithms stands at 71%, while the F1 score is marginally higher for LightGBM at 70% versus 67% for Logistic Regression. Random Forest also shows strong performance with an accuracy of 70%, precision at 65%, and both recall and F1 scores at 70% and 67%, respectively. Algorithms such as the K-Nearest Neighbors and Naive Bayes offer moderate performance, with accuracy figures of 68% and 64%, precision at 60% and 62%, recall at 63% and 64%, as well as F1 scores of 63% and 62%, in that order. The XGBoost and CatBoost algorithms exhibit identical accuracy at 70%. XGBoost has somewhat higher precision at 68% compared to CatBoost's 65%, recall for both is at 70%, and their F1 scores are identical at 68%. The Decision Tree algorithm presents a lower accuracy of 57%, with a precision of 62%, a recall of 57%, and an F1 score of 59%, showing it may not be as effective as the other algorithms in this context.Finally, the Artificial Neural Network and AdaBoost algorithms show the least effective performances with this dataset. The Artificial Neural Network has an accuracy of 50%, precision of 62%, recall of 59%, and an F1 score of 60%. AdaBoost notably has the lowest accuracy at 41%, though its precision is higher at 69%, with both recall and F1 score at 41% and 48%, respectively.It is crucial, however, to consider these performance metrics in context with the specific application and data sensitivity to each metric for a complete assessment of classifier effectiveness.

## 4.2 Accuracy of Models Considering Feature Importance

| Algorithms | New | Previous | Change |
|---|---|---|---|
| Logistic Regression | 71.00% | 71.00% | 0.00% |
| Decision Tree | 57.00% | 54.00% | +3.00% |
| K-Nearest Neighbors | 68.00% | 68.00% | 0.00% |
| Naive Bayes | 64.00% | 64.00% | 0.00% |
| Random Forest | 70.00% | 68.00% | +2.00% |
| SVM | 71.00% | 71.00% | +3.00% |
| Gradient Boosting | 74.00% | 72.00% | -1.00% |
| XGBoost | 70.00% | 67.00% | +3.00% |
| AdaBoost | 41.00% | 42.00% | -1.00% |
| LightGBM | 71.00% | 68.00% | +3.00% |
| CatBoost | 72.00% | 72.00% | 0.00% |
| Artificial Neural Network | 50.00% | 66.00% | -16.00% |

**Table 3** Comparison of New and Previous Accuracy

This section evaluates the influence of feature selection on the performance of various machine learning algorithms. By identifying and utilizing seven attributes with high feature importance scores, we retrained and tested our models to observe any variances in accuracy. Contrasted with the accuracy, our findings reveal that some algorithms benefit from feature selection, while others do not. The New Accuracy column represents the accuracy after implementing feature importance. The application of feature selection improved accuracy for algorithms such as Decision Tree, Random Forest, Gradient Boosting, XGBoost, and LightGBM. These models showed a noticeable increase in accuracy, highlighting the role played by feature selection in enhancing predictive performance. For instance, the Decision Tree demonstrated a significant improvement, with a 3% increase in accuracy, implying a better fit for the reduced feature set. Conversely, other models like AdaBoost and the Artificial Neural Network exhibited a decrease in performance when subjected to feature importance, with the Artificial Neural Network showing a substantial reduction in accuracy. This suggests that certain algorithms may depend on a broader set of features for optimal performance or that the feature selection process may have excluded critical information necessary for these models. In conclusion, feature importance has proven to be beneficial for several algorithms in increasing predictive accuracy.

## 4.3 Performance based on 20-fold cross-validation

| Predicted Model | Accuracy |
|---|---|
| Logistic Regression | 0.6425 |
| Support Vector Machine | 0.6150 |
| Decision Tree | 0.5980 |
| Random Forest | 0.6610 |
| K-Nearest Neighbour | 0.6410 |
| Naive Bayes | 0.6370 |
| Gradient Boosting | 0.6980 |
| XGBoost | 0.6780 |
| AdaBoost | 0.2770 |
| LightGBM | 0.7065 |
| CatBoost | 0.6780 |
| Artificial Neural Network | 0.6210 |

**Table 4** Classifiers Performance based on 20- fold Cross-validation

Stratified K-Fold is better than K-Fold as it tries to balance different data points or include all types in each fold, thereby ensuring that the model is trained and evaluated on representative samples from each class. So here we use Stratified K-Fold. In 20-fold cross-validation, we use 95% of data to train the classifier and 5% of data to test the classifier. Table 4 showcases the outcomes of 20-fold cross-validation for 12 different classifiers.

Again, In 10-fold cross-validation, we use 90% of data to train the classifier and 10% of data to test the classifier.
Table 5 showcases the outcomes of 10-fold cross-validation for 12 different classifiers.

## 4.4 ROC curve(AUC)

ROC curve means Receiver Operating Characteristic curve which is a plot of True Positive Rate versus False Positive Rate. ROC curve is usually used for comparison of multiple classifiers. Here we are using the ROC curve for expressing Classifier performance. The ROC curve displayed in Figure 2 highlights the exceptional classification performance achieved.

| Predicted Model | Accuracy |
|---|---|
| Logistic Regression | 0.624 |
| Support Vector Machine | 0.612 |
| Decision Tree | 0.571 |
| Random Forest | 0.659 |
| K-Nearest Neighbour | 0.632 |
| Naive Bayes | 0.636 |
| Gradient Boosting | 0.687 |
| XGBoost | 0.674 |
| AdaBoost | 0.352 |
| LightGBM | 0.699 |
| CatBoost | 0.674 |
| Artificial Neural Network | 0.614 |

**Table 5** Classifiers Performance on 10-fold Cross-validation

| Classifier | AUC Score |
|---|---|
| Logistic Regression | 0.80 |
| Support Vector Machine | 0.81 |
| Decision Tree | 0.70 |
| Random Forest | 0.78 |
| K-Nearest Neighbour | 0.77 |
| Naive Bayes | 0.80 |
| Gradient Boosting | 0.82 |
| XGBoost | 0.79 |
| AdaBoost | 0.68 |
| LightGBM | 0.81 |
| CatBoost | 0.82 |
| Artificial Neural Network | 0.75 |

**Table 6** AUC Scores for Different Classifiers

As shown in Figure 2 and Table 6, it was found that Gradient Boosting and CatBoosting achieved an AUC of 0.82 which is the highest AUC score. It is sufficiently good. The second top-performing algorithms were Support Vector Machine and LightGBM. The performance of Logistic Regression and Naive Bayes is also good. In summary, after evaluating 12 algorithms, it was found that the Gradient Boosting and the CatBoosting demonstrated the highest level of performance among them all.

## 4.5 Hyperparameter Tuning

Hyperparameters are parameters whose values control the learning process. These parameters are employed to achieve the efficiency of different models. RandomizedSearchCV with 20 fold was applied while tuning the parameters of different classifiers as they randomly chose different combinations other than brute force techniques. The best parameters of different models are listed in Table 8.
At first, we used a method called random search CV with 20-cross validation. This means we tried out lots of different combinations of settings for each model. Then, by splitting the data into 20 parts and testing each combination, we found the settings that gave the highest accuracy. Once we found these best settings, we plugged them back into each model. As a result, the models performed even better, giving us more accurate results than before.

| Model | Score(20-CV) | Best Parameter |
|---|---|---|
| Logistic Regression | 0.6515 | 'solver': 'newton-cg', 'multi_class': 'ovr', 'C': 10 |
| Support Vector Machine | 0.6865 | 'kernel': 'linear', 'C': 1.0 |
| Decision Tree | 0.6223 | 'splitter': 'random', 'min_samples_split': 2, 'min_samples_leaf': 2, 'max_features': None, 'max_depth': 19, 'criterion': 'gini' |
| Random Forest | 0.6851 | 'n_estimators': 200, 'max_depth': None, 'criterion': 'entropy' |
| Naive Bayes | 0.6341 | - |
| KNN | 0.6414 | 'weights': 'distance', 'n_neighbors': 8, 'leaf_size': 5, 'algorithm': 'auto' |
| Gradient Boosting Machine | 0.7070 | 'n_estimators': 50, 'max_depth': 4, 'learning_rate': 0.1 |
| AdaBoost | 0.5540 | 'n_estimators': 50, 'learning_rate': 0.01 |
| XGBoost | 0.7085 | 'n_estimators': 100, 'min_child_weight': 0, 'max_depth': 4, 'learning_rate': 0.15 |
| LightGBM | 0.7100 | 'num_leaves': 7, 'n_estimators': 100, 'min_data_in_leaf': 10, 'max_depth': 3, 'learning_rate': 0.05 |
| CatBoost | 0.71580 | 'learning_rate': 0.1, 'iterations': 150, 'depth': 5 |
| Artificial Neural Network | 0.6800 | 'solver': 'adam', 'max_iter': 500, 'learning_rate': 'constant', 'hidden_layer_sizes': (50, 50), 'alpha': 0.001, 'activation': 'logistic' |

**Table 7** Best Parameters for Different Models

| Model | Score |
| --- | --- |
| Logistic Regression | 0.69 |
| Support Vector Machine | 0.68 |
| Decision Tree | 0.59 |
| Random Forest | 0.75 |
| Naive Bayes | 0.67 |
| K-Nearest Neighbour | 0.64 |
| Gradient Boosting | 0.75 |
| AdaBoost | 0.62 |
| XGBoost | 0.75 |
| LightGBM | 0.76 |
| CatBoost | 0.75 |
| Artificial Neural Network | 0.70 |

**Table 8** Splitting Datasets Score with Best Parameter

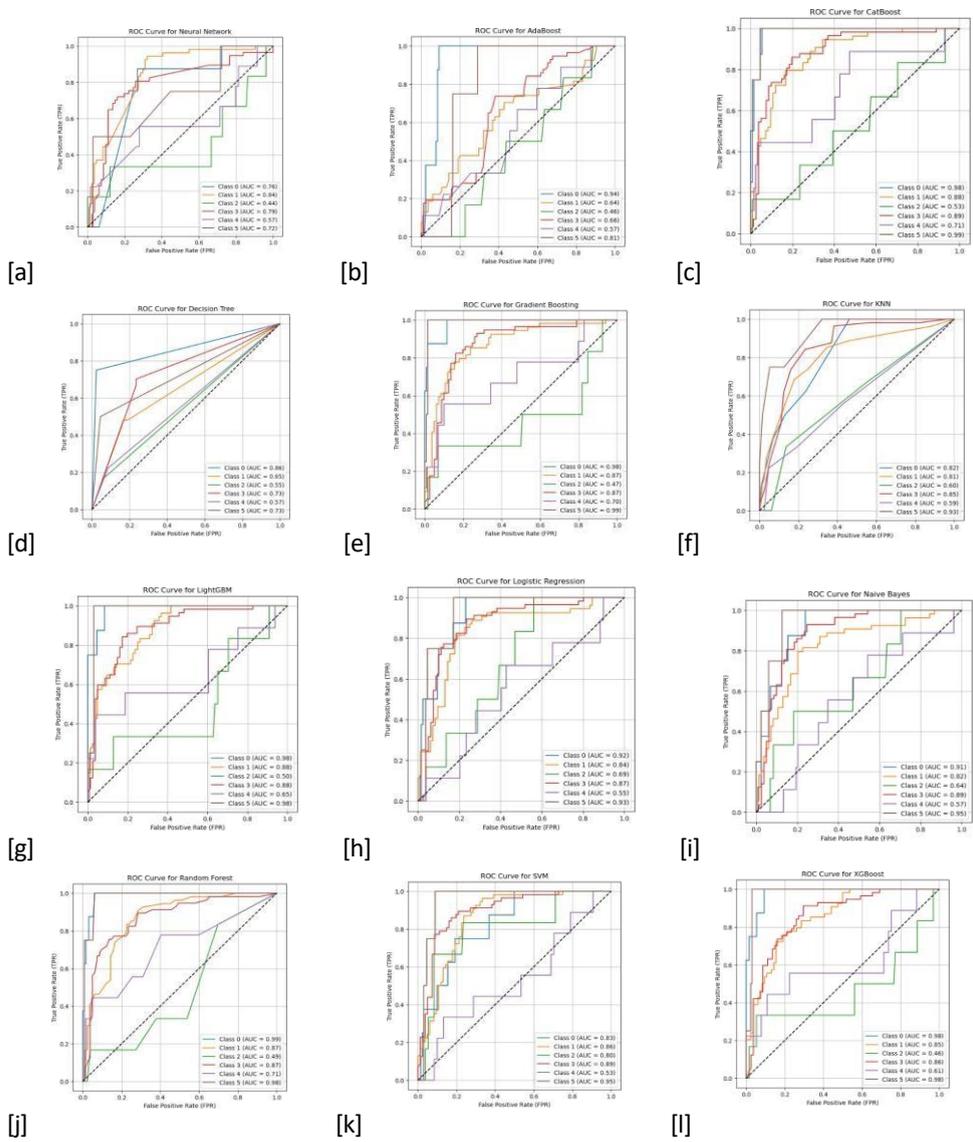

**Fig. 2** ROC curve(AUC)

Figure 2 : The ROC curves with AUC for different algorithms (a) Neural Network (b) AdaBoost (c) CatBoost (d) Decision Tree (e) Gradient Boosting (f) K-Nearest Neighbour (g) LightGBM (h) Logistic Regression (i) Naive Bayes (j) Random Forest (k) Support Vector Machine (l) XGBoost

Out of twelve models with their best parameters, six models perform well enough. At this point, we introduce the Voting Classifier as an optimizer. The Voting classifier takes all of these models as base learners and applies the majority voting concept to select majority-voted performance among all models. The model achieves 77% accuracy which serves as the final prediction of cancer treatment.

| Algorithm | Score | Parameters |
| --- | --- | --- |
| Logistic Regression | 0.69 | 'solver': 'newton-cg', 'multi_class': 'ovr', 'C': 10 |
| Support Vector Machine | 0.68 | 'kernel': 'linear', 'C': 1.0 |
| Decision Tree | 0.59 | 'splitter': 'random', 'min_samples_split': 2, 'min_samples_leaf': 2, 'max_features': None, 'max_depth': 19, 'criterion': 'gini' |
| Random Forest | 0.75 | 'n_estimators': 200, 'max_depth': None, 'criterion': 'entropy' |
| Naive Bayes | 0.67 | - |
| K-Nearest Neighbors | 0.64 | 'weights': 'distance', 'n_neighbors': 8, 'leaf_size': 5, 'algorithm': 'auto' |
| Gradient Boosting Machine | 0.75 | 'n_estimators': 50, 'max_depth': 4, 'learning rate': 0.1 |
| AdaBoost | 0.62 | 'n_estimators': 50, 'learning rate': 0.01 |
| XGBoost | 0.75 | 'n_estimators': 100, 'min_child_weight': 0, 'max_depth': 4, 'learning rate': 0.15 |
| LightGBM | 0.76 | 'num_leaves': 7, 'n_estimators': 100, 'min_data_in_leaf': 10, 'max_depth': 3, 'learning rate': 0.05 |
| CatBoost | 0.75 | 'learning_rate': 0.1, 'iterations': 150, 'depth': 5 |
| Artificial Neural Network | 0.70 | 'solver': 'adam', 'max_iter': 500, 'learning_rate': 'constant', 'hidden_layer_sizes': (50, 50), 'alpha': 0.001, 'activation': 'logistic' |

**Table 9** Model Performance and Parameters

## 5 Conclusion

Medical data can be analyzed using a variety of data mining and machine-learning techniques. The biggest obstacle was in the fields of machine learning, data cleansing, and data mining, where we had to develop precise and computationally effective classifiers for medical applications. Logistics Regression, Support Vector Machine,

Decision Tree, Random Forest, K-Nearest Neighbours, Naive Bayes, Gradient Boosting, XGBoost, AdaBoost, LightGBM, CatBoost, and ANN are the twelve machine learning methods that we have examined. Among the twelve machine learning algorithms, Random Forest yields the highest accuracy, at 70%, while Decision Tree yields the lowest accuracy, at 52% when considering all features. However, after eliminating the superfluous features, Gradient Boosting produced the best accuracy (74%), while AdaBoost produced the lowest accuracy (42%). We assessed the models using separate test data sets and various cross-validations. We have made use of real cancer patient data that we have acquired from hospitals. We had to convert the prescription data into digital format to work with the data. Due to the low volume of the data, the accuracy of the model suffered. To improve our efficiency while gathering data, assessing the models, and creating reliable machine learning-based prediction models, we will do additional research in the future and collect a large amount of data from different entities.

## 6 Declarations

- Conflict of Interest: The authors have no conflicts of interest to disclose.
- Ethics Approval and Consent to Participate: Before collecting the dataset, ethical approval was obtained from the organization, and an agreement was made with the participants. This protocol was followed throughout the study.
- Data Availability: According to the codes of conduct, the data is not available online but will be provided upon request.
- Code Availability: The code used in this study will be provided upon request from the authors.